\documentclass{emulateapj}
\usepackage{graphicx}
\usepackage{mathrsfs}	
\usepackage{amsmath}
\usepackage{subfigure}
\usepackage{color}

\shorttitle{Velocity-Dispersion Evolution}
\shortauthors{Shu et al.}
 
\begin{document}
 
\title{Evolution of the Velocity-Dispersion Function of Luminous Red Galaxies: \\ A Hierarchical Bayesian Measurement}

\author{\mbox{Yiping Shu\altaffilmark{1}}}
\author{\mbox{Adam S. Bolton\altaffilmark{1}}}
\author{\mbox{David J. Schlegel\altaffilmark{2}}}
\author{\mbox{Kyle S. Dawson\altaffilmark{1}}}
\author{\mbox{David A. Wake\altaffilmark{3}}}
\author{\mbox{Joel R. Brownstein\altaffilmark{1}}}
\author{\mbox{Jon Brinkmann\altaffilmark{4}}}
\author{\mbox{Benjamin A. Weaver\altaffilmark{5}}}

\altaffiltext{1}{Department of Physics and Astronomy, University of Utah,
115 South 1400 East, Salt Lake City, UT 84112, USA ({\tt yiping.shu@utah.edu, bolton@astro.utah.edu})}
\altaffiltext{2}{Lawrence Berkeley National Laboratory, Berkeley, CA 94720, USA}
\altaffiltext{3}{Department of Astronomy, Yale University, New Haven, CT 06520 USA}
\altaffiltext{4}{Apache Point Observatory, Apache Point Road, P.O. Box 59, 
Sunspot, NM 88349, USA}
\altaffiltext{5}{Center for Cosmology and Particle Physics, New York University, 
NY 10003, USA}

\begin{abstract}
We present a hierarchical Bayesian determination of the velocity-dispersion function of
approximately 430,000 massive luminous red galaxies(LRGs) observed at relatively low
spectroscopic signal-to-noise ratio (SNR $\sim$3--5 per 69\,km\,s$^{-1}$) by the
Baryon Oscillation Spectroscopic Survey (BOSS) of the Sloan Digital Sky Survey III (SDSS-III)\@.
We marginalize over spectroscopic redshift errors, and use the full velocity-dispersion
likelihood function for each galaxy to make a self-consistent determination of
the velocity-dispersion distribution parameters as a function of absolute magnitude
and redshift, correcting as well for the effects
of broadband magnitude errors on our binning.
Parameterizing the distribution at each point in the luminosity--redshift
plane with a log-normal form,
we detect significant evolution in
the width of
the distribution toward higher intrinsic scatter at higher redshifts.
Using a subset of deep re-observations of BOSS galaxies, we demonstrate
that our distribution-parameter estimates are unbiased regardless of spectroscopic SNR\@.
We also show through simulation that our method introduces no systematic
parameter bias with redshift.
We highlight the advantage of the hierarchical Bayesian method
over frequentist ``stacking'' of spectra, and illustrate how our
measured distribution parameters can be adopted as informative priors for
velocity-dispersion measurements from individual noisy spectra.

\end{abstract}


\slugcomment{to be submitted to The Astronomical Journal}

\maketitle

\section{Introduction}

Massive elliptical galaxies \citep[EGs:][]{Hubble} are one of the most
important classes of astrophysical objects for galaxy evolution and
cosmology.  They represent the end stage of hierarchical galaxy-formation
processes \citep[e.g.,][]{kauffmann_93, baugh_96}, and therefore their properties and scaling relations
represent a key test for theories of galaxy formation within
a cosmological context.  In addition, since they are the most
luminous and highly clustered galaxies, they serve as ideal
cosmological tracers of clusters and large-scale structure \citep[e.g.,][]{eisenstein_05}.


To a first approximation,
EGs are ``pressure-supported'' rather than rotationally supported
\citep[e.g.,][]{bertola_75, illingworth_77, binney_78},
with their stellar motions characterized by a velocity dispersion $\sigma$.
Among the many observational parameters of massive elliptical galaxies,
${\sigma}$ is unique in its direct sensitivity
to the depth of the galaxy's gravitational potential (and therefore to its mass),
and in its relatively weak dependence on observational aperture.
In combination with
galaxy sizes (i.e., half-light radii), velocity dispersions
can be used to determine ``dynamical masses'' that are independent of stellar-population
assumptions \citep[e.g.,][]{padmanabhan_04, bolton_08b}.
Dynamical masses can then in turn be used to trace the evolution
of EGs at fixed mass \citep[e.g.,][]{vdm_07, vdw_08, cappellari_09},
indicating a nuanced dynamical history
despite generally passive star-formation
histories at $z < 1$ \citep[e.g.,][]{thomas_05,cool_08}.
Stellar velocity dispersion is also the most important single predictor
of strong gravitational lensing cross sections \citep[e.g.,][]{Turner, bolton_08a},
and can be used in combination with
strong lensing observations to constrain the central mass-density structure of elliptical
galaxies at cosmological distances \citep[e.g.,][]{kt02,tk04,koopmans_06}.
Stellar velocity dispersions are tied to nearly all other properties
of EGs through multiple empirical scaling relations.
\citet{FJR} found a correlation between luminosities
of early-type galaxies and their velocity dispersions ${\sigma}$ known as the
Faber-Jackson Relation (FJR).   The relation of \citet{Kormendy} ties the surface
brightness $\langle \rm I \rangle _e$ with the effective radius $\rm R_e$.
Both the FJR and Kormendy relations can be viewed as projections of the
``fundamental plane'' \citep[FP, e.g.,][]{Djorgovski, Dressler, Bernardi03III}
within the space spanned by $\rm \log_{10} R_e$, $\rm \langle I \rangle _e$ and
$\rm \log_{10}{\sigma}$.
Furthermore, central black hole mass has been found to be correlated with
the velocity dispersion of the bulge via the $\rm M_{BH} - {\sigma}$
relation\citep[e.g.,][]{Ferrarese, Gebhardt, Kormendy09}.
Together, these relations provide multiple constraints on
the structure, formation, and evolution of EGs \@.


Although velocity dispersion plays a starring role in the study of
EGs, it is an ``expensive'' observable that must be measured spectroscopically.
Hence, large samples of galaxies with well-measured velocity dispersions across cosmic time are largely
unavailable.  Measurements of $\sigma$ are made by quantifying the line-of-sight Doppler
broadening of absorption lines relative to a set of template stellar spectra,
either via the Fourier method \citep[e.g.,][]{Sargent, Tonry} or the direct-fitting
method \citep[e.g.,][]{Burbidge, Rix}. Both methods rely on the quality of galaxy spectra:
for spectra of low signal-to-noise ratio (SNR), uncertainties in the measured stellar
velocity dispersion can be very large and significantly non-Gaussian.  This aspect is of
particular concern for galaxies at cosmological distance (faint even if luminous),
which can only be measured at high SNR through substantial investment
of spectroscopic observing time and aperture.

In this paper, we introduce a hierarchical Bayesian statistical method
to measure the parameters of the distribution of stellar velocity dispersions
within a population of galaxies that has been observed with relatively low
spectroscopic signal-to-noise ratio.  We apply the method to approximately
430,000 luminous red galaxy (LRG) targets from the Baryon Oscillation
Spectroscopic Survey \citep[BOSS:][]{schlegel_09}, one of four survey
projects within the Sloan Digital Sky Survey III
\citep[SDSS-III:][]{sdss3}.  We quantify the evolution of the
velocity-dispersion function of BOSS galaxies, and detect significant evolution in the
intrinsic population RMS of $\log_{10} \sigma$ at fixed absolute magnitude since $z \approx 0.8$. 

This paper is organized as follows. In Section~\ref{sec:data},
we describe the sample selection
and the method for velocity dispersion measurement.  Section~\ref{sec:stats}
presents our statistical method for the measurement of the distribution of stellar
velocity dispersions within a population of galaxies, including
a verification using high-SNR re-observations of a sub-sample of galaxies and 
a test for redshift-dependent systematic biases.
Section~\ref{sec:evol} presents the results of our
application of this method to the BOSS sample, showing the
evolution of the velocity-dispersion function at fixed magnitude.
Discussion and conclusions are presented in Section~\ref{sec:disco}.
Throughout the paper, we assume a standard general-relativistic cosmology with
${\Omega}_{\rm m} = 0.3$, ${\Omega}_{\rm \Lambda} = 0.7$ and
$\rm H_{\rm 0}=70 \rm \ km \rm \ s^{-1} \rm \ Mpc^{-1}$. 

\section{Spectroscopic Data}
\label{sec:data}

\subsection{Sample Selection}

We use spectroscopic data obtained by the BOSS project via the 2.5-m SDSS telescope located
at Apache Point Observatory in Sunspot, New Mexico \citep{Telescope}.
The primary science goal of BOSS is the detection of the baryon
acoustic feature in the two-point correlation function of galaxies (and quasar absorption systems),
from which to constrain the distance--redshift relation and the nature of dark energy.
BOSS also offers
a unique resource for the study of the properties and evolution of massive galaxies.
The BOSS footprint covers approximately 10,000 deg$^2$ in five imaging filters \citep[\emph{ugriz},][]{Fukugita},
and will by 2014 obtain spectra of about 1.5 million LRGs out to redshift
$\rm z \simeq 0.8$.
Note that the majority of the BOSS LRG targets are massive EGs, although there is a small fraction of late-type 
galaxies as well as unresolved multiples, particularly at the higher redshift end \citep{Masters11}.

The BOSS spectra are broadly comparable to \mbox{SDSS-I} \citep{SDSSI} spectra in resolution ($R \approx 2000$),
and cover a wavelength range from 3,600\AA\ to 10,000\AA\@.
The primary design goal of BOSS is to measure as many redshifts as efficiently as possible,
in order to map the largest possible volume of the universe.
Consequently, the SNR of the galaxy spectra is significantly lower than in SDSS-I,
with typical SNR values of 3 to 5 per 69\,km\,s$^{-1}$ (rebinned) pixel,
as compared with $\gtrsim$ 10 per pixel in SDSS-I\@.
Thus although the BOSS spectroscopic database is by far the largest
available for the study of massive galaxies, the individual spectra
are well below the SNR threshold of about 10 per \AA\ generally regarded as a minimum
for acceptable velocity-dispersion measurement on a galaxy-by-galaxy basis.
Motivated by this context, we develop the Bayesian analysis method presented below.

Spectroscopic calibration, extraction, classification, and redshift
measurement of all BOSS galaxy spectra
are carried out using the \texttt{idlspec2d} software \citep[see, e.g.,][]{DR8},
written originally for SDSS-I and recently updated to handle the
data format and noise regime of BOSS\@.
In selecting our analysis sample, we make the following cuts
based upon the redshift pipeline output:
\begin{itemize}
\item We use only the best spectroscopic observation of any given galaxy target as some objects are observed more than
  once (\texttt{SPECPRIMARY}=1 according to SDSS terminology).
\item We use only objects that were both targeted as galaxies and spectroscopically confirmed
 as galaxies 
\item We require a confident redshift measurement with no warning flags 
(\texttt{ZWARNING} = 0 according to SDSS terminology).\footnote{For BOSS galaxies,
the specifically relevant flag is \texttt{ZWARNING\_NOQSO} = 0 (Bolton et al., in prep.)}
\end{itemize}
These cuts return approximately 430,000 galaxies from the first 1.5 years of BOSS
spectroscopic observations, with redshifts ranging from zero to 1, but
concentrated primarily over the interval $0.2 < z < 0.8$.

For all selected galaxies, we use the measured spectroscopic redshifts
and SDSS broadband imaging colors to compute absolute rest-frame $V$-band magnitudes
and associated uncertainties via
the \texttt{sdss2bessell} routine implemented in the \texttt{kcorrect} software of
\citet{kcorrect}.



The details of the BOSS galaxy target selection, and the corresponding
incompletenesses, are the subject of a
separate paper (Padmanabhan et al., in preparation).  Here we summarize the target selection cuts for the two main
galaxy target classes that we focus upon in our current study.
The first is the CMASS sample (for ``constant mass''), which is selected photometrically
to deliver LRGs of approximately constant stellar mass over the redshift interval $0.3 < z < 0.8$,
and which constitutes approximately 76\% of the galaxies selected above.
The second sample, LOZ, is selected to target LRGs at lower redshifts, and represents
the remaining 24\% of the selected galaxies.
Defining the following quantities \citep{Eisenstein01,Cannon06}:
\begin{eqnarray}
  c_{\parallel} &~=~& 0.7(g-r)+1.2[(r-i)-0.18]  \\
  c_{\perp} &~=~& (r-i)-(g-r)/4.0-0.18  \\
  d_{\perp} &~=~& (r-i)-(g-r)/8.0 \\
  \mbox{ifiber2} &~=~&  \mbox{$i$-band fiber magnitude for 2$''$ fibers,}
\end{eqnarray} 
the CMASS sample is defined by the photometric cuts:
\begin{eqnarray}
  & 17.5<i<19.9 \label{cmass1} \\
  & r-i <2  \label{cmass2} \\
  & d_{\perp} > 0.55 \label{cmass3} \\
  & {\rm ifiber2} < 21.7 \label{cmass4} \\
  & i < 19.86+1.60 (d_{\perp}-0.80) \label{cmass5}
\end{eqnarray}
as well as a cut to exclude galaxies with major-axis
half-light radii greater than 8$^{\prime \prime}$.
Equations~(\ref{cmass1}) and (\ref{cmass3}) aim to select galaxies between
redshifts $z\sim0.4$--0.8, while Equation~(\ref{cmass5})
attempts to impose a cut at constant stellar mass across this redshift range.
The LOZ sample is defined by the cuts:
\begin{eqnarray}
  & r < 13.5 + c_{\parallel}/0.3 \label{loz1} \\
  & |c_{\perp}| < 0.2 \label{loz2} \\
  & 16< r <19.6 \label{loz3} ~.
\end{eqnarray}
Equation~(\ref{loz1}) sets up a magnitude threshold as a function of redshift and
Equation~(\ref{loz2}) picks out low-redshift galaxies specifically.

\begin{figure}[t]
\centering
\includegraphics[width=0.5\textwidth]{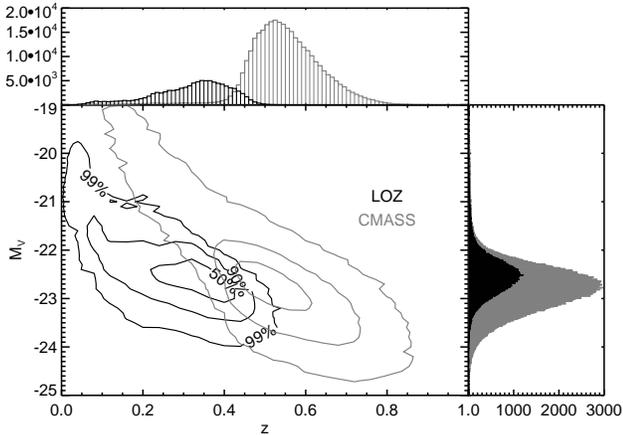}
\caption{\label{fig:gal_dist}
Distribution of galaxies for our sample, along with histograms of redshift $z$ and V-band absolute magnitudes $M_V$.
LOZ galaxies \textit{(black)} and CMASS galaxies \textit{(gray)} are plotted separately.  For both the samples,
contours are drawn at constant number density in the $z$--$M_V$ plane, enclosing 50\%, 90\%, and 99\% of the sample.}
\end{figure}

The redshift--absolute magnitude distributions of these two BOSS galaxy
samples, with associated 1D projections,
are plotted in Figure~\ref{fig:gal_dist}.
In the following analysis, we will treat the two populations separately, since the combined sample
does not define a simple locus in luminosity--redshift space, with LOZ galaxies being of
generally higher luminosity over the redshift range where the two samples overlap.

\subsection{Velocity Dispersion Extraction}

Our strategy for extracting velocity-dispersion information is to make use of the full velocity-dispersion
likelihood function for each galaxy spectrum.
To do this, we make use of the IDL routine \texttt{vdispfit} within the
\texttt{idlspec2d} product of spectroscopic analysis software.  This software has been extensively
tested in the SDSS-I, and has been upgraded for the analysis of BOSS data.  Velocity
dispersions measured with this software have been the basis for multiple studies
of the dynamics of EGs \citep[e.g.,][]{Bernardi03I,Bernardi03II,Bernardi03III,sheth_03,padmanabhan_04,koopmans_06}.
To summarize briefly: \texttt{vdispfit} uses a set of stellar eigenspectra derived from a principal-component analysis (PCA) decomposition of
the ELODIE stellar spectrum library \citep{ELODIE}.
The eigenspectra are convolved and binned to the resolution and sampling of the BOSS spectra,
then broadened by Gaussian kernels of different trial velocity dispersions.
The broadened templates are then shifted to the redshift of the galaxy under consideration. After masking out regions
containing common emission lines, a linear least-squares fit is performed to obtain a best-fit model spectrum
at each trial velocity dispersion.  The resulting curve of $\chi^2$ as a function of trial velocity dispersion
encodes the likelihood function of velocity dispersion given the data.  For measurements from
high signal-to-noise spectra, the position of the minimum $\chi^2$ is adopted as the maximum-likelihood
estimate of the galaxy's velocity dispersion.  Below, rather than adopt these estimates, we will
work with the full likelihood function.

In this procedure, we must choose the number of stellar eigenspectra to use in forming the template basis.
The pipeline analysis of SDSS-I data used the first 24 PCA modes.  For the much lower signal-to-noise BOSS
data, an acceptable $\chi^2$ can be obtained using only the first 5 PCA modes,
and hence we restrict our basis to this smaller number of eigenspectra so as to avoid
fitting noise fluctuations.

As described above, before being fit to the galaxy spectra,
the stellar eigenspectra are shifted by the appropriate galaxy redshifts.
If the redshifts have non-negligible errors, the corresponding offsets can introduce a bias into the measured velocity dispersion.
Although the BOSS spectra provide redshifts with a precision well in excess of what is required
for large-scale structure studies and absolute-magnitude determinations,
their errors can be non-negligible on the scale of internal galaxy velocity dispersions.
Therefore, we implement a marginalization over redshift errors in our analysis.
Specifically, we modify the \texttt{vdispfit} routine to take a radial velocity-marginalization range $\Delta z$
(expressed in constant-velocity pixels)
and the redshift error $\delta z$ (the $\pm 68$\% confidence interval as estimated by the \texttt{idlspec2d} pipeline) as arguments. Then we calculate $\chi^2(\sigma, z)$
for a set of trial redshifts in the range $z \pm \Delta z$ and
define a new effective $\chi^2(\sigma)$ by integrating over $z$ as
\begin{equation}
  \chi^2(\sigma) = -2 \ln \left( \int_{z - \Delta z} ^{z + \Delta z} \! dz \,
    \exp\left[-\frac{{\chi}^2(\sigma, z)}{2} \right] \,  p(z) \right)~,
\end{equation}

We assume a Gaussian probability distribution for z given by 
\begin{equation}
  p (z) \propto \exp \left[ -\frac{(z-z_{\mathrm{best}})^2}{2 \delta z^2} \right]
\end{equation}
where $z_{\mathrm{best}}$ is the best-estimate redshift from the BOSS spectroscopic pipeline.
The choice of a Gaussian prior is made because
the galaxy redshifts have been determined using absorption and emission-line information
over the full optical range of the BOSS spectrograph, whereas the velocity-dispersion
fitting is done only over the 4100--6800\,\AA\ rest-frame range covered by the ELODIE spectra,
while also masking the wavelength positions of common emission lines.
We also explored the use of a flat prior to completely marginalize over redshift
in the velocity dispersion analysis, and found only a negligible change
(at most a few percent) in the derived relations.
For most galaxies, the effect of this marginalization on the $\chi^2$ curve
is insignificant, but since we wish to avoid introducing any spurious broadening
into our population analysis, we apply the procedure to all spectra.

In this work, we do not make any aperture correction for velocity dispersions, although the angular BOSS fiber radius of
1$^{\prime \prime}$ subtends a different physical length scale as a function of redshift. Since aperture velocity dispersions
are seen in the local universe to depend on aperture radius only to a weak power of approximately 0.04 to 0.06
\citep[e.g.,][]{jorgensen_vdisp, mehlert_03, capp_sauron}, this effect should be relatively insignificant. 
For example, taking a redshift range spanning
the majority of our CMASS sample, the angular size of a fixed physical length at $z=0.8$
is about 72\% of its angular size at $z = 0.4$.  Assuming the velocity dispersion within an aperture
decreases as the aperture to the power $-0.05$ (a representative compromise value between the previous three references),
this would correspond to a systematic change in measured velocity dispersion of about 1.7\%,
which is well below the level of other uncertainties in our analysis.
In addition, the typical atmospheric seeing of approximately 1.8$^{\prime \prime}$ delivered to the BOSS spectroscopic
focal plane will dilute the significance of the varying projected fiber scale.  Essentially, BOSS velocity
dispersions will represent a fair luminosity-weighted average value over the half-light radius of
most target galaxies, which have half-light radii on the order of 1$^{\prime \prime}$.


\section{Statistical Population Analysis Formalism}
\label{sec:stats}

The results of \citet{Bernardi03II} suggest that the distribution of
velocity dispersions for early-type galaxies at fixed luminosity
can be well approximated by a log-normal function.  Motivated by this,
we will assume a Gaussian distribution in $\log_{10}{\sigma}$ with mean $m$ and intrinsic scatter $s$:
\begin{equation}
\label{eq:pdf}
   p (\log_{10}{\sigma} | m, s) = \frac{1}{\sqrt{2\pi}s} \exp \left[
{{-(\log_{10}{\sigma}-m)^2} \over {2 s^2}} \right]
\end{equation}
We will treat $m$ and $s$ as functions of redshift and absolute magnitude, although we will suppress this
dependence in our notation for convenience.  Compared to the SDSS-I studies by \citet{Bernardi03II}
and \citet{sheth_03}, we have 
a much larger sample with greater redshift coverage, so we may investigate the evolution of both the mean
and intrinsic scatter of $\log_{10} {\sigma}$ with redshift and luminosity as encoded by these two distribution parameters (See also \citealt{Bezanson11} for a complementary analysis in terms
of photometric velocity-dispersion proxies). 
Our strategy will be to analyze samples binned by an interval of 0.04 in redshift $z$,
and by 0.1 in absolute magnitude $M_V$.

\subsection{Frequentist Approach}

\begin{figure}[t]
\centering
  \includegraphics[width=0.5\textwidth]{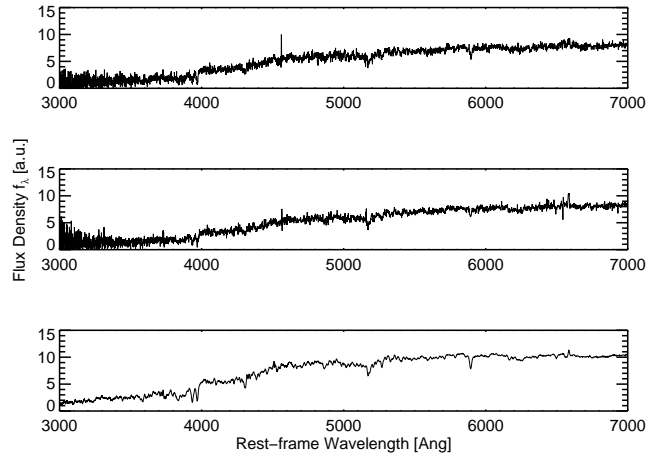}
\caption{\label{fig:stack}
Results for stacking of spectra within a single redshift-luminosity bin. The top two panels show typical individual spectra, while the bottom panel shows the high-SNR stacked spectrum for that bin, resulting from averaging the spectra of $\sim 200$ galaxies.}
\end{figure}

As mentioned above, the SNR of BOSS galaxy spectroscopy is typically rather low, especially at the high-redshift end of the survey.
Therefore, point estimation of the velocity dispersion of individual galaxies is of
questionable reliability.
Hence, we resort to analyzing the data by binning galaxies in the $z$--$M_V$ plane,
requiring at least 100 galaxies in every single bin.
The most obvious first approach to determining the mean velocity dispersion in these bins
is to remove the small relative redshift
differences within the bin, stack all the spectra directly, and analyze the resulting high-SNR combination
(see Figure~\ref{fig:stack}).
Although we do not adopt this method for our ultimate determinations of $m$ and $s$, it is instructive to
consider how such an approach relates to these parameters.

While a velocity dispersion can be measured at high SNR from the stacked spectrum,
the measured value bears a non-trivial relation to the parameters $m$ and $s$, which we now derive.
Assuming equal luminosities within the bin (which basically holds by construction due to binning in absolute magnitude),
what we measure from the stack ${\sigma}_{\rm stack}^2$ is the population-weighted expectation value of ${\sigma}^2$, i.e.
\begin{equation}
  {\sigma}_{\rm stack}^2 = \langle {\sigma}^2 \rangle = \int \! {\sigma}^2 \, p(\log_{10}{\sigma} | m, s) \, \mathrm{d}\log_{10} {\sigma}~.
\end{equation} 
The variance of ${\sigma}_{\rm stack}^2$ is given by
\begin{equation}
  \mathrm{Var}({\sigma}_{\rm stack}^2) = \frac{1}{N} \mathrm{Var}({\sigma}^2) = 
  \frac{1}{N} (\langle {\sigma}^4 \rangle - \langle {\sigma}^2 \rangle^2)
\end{equation}
with $N$ being the number of galaxies in the bin.

Making use of the following relation, which can be derived for our log-normal form of Equation~(\ref{eq:pdf}):
\begin{eqnarray}
\label{eq:sigma}
  \langle {\sigma}^n \rangle &=& \int \! {\sigma}^n \, p(\log_{10}{\sigma} | m, s) \, \mathrm{d}\log_{10} {\sigma} \nonumber \\
                             &=& 10^{[n\,m+n^2\,\ln(10)\,s^2/2]}~,
\end{eqnarray}
we find that
\begin{eqnarray}
  {\sigma}_{\rm stack}^2 &=& 10^{[2\,m+2\,\ln(10)\,s^2]} \label{eq:sigma_stack} \\
  Var({\sigma}_{\rm stack}^2) &=& \frac{({\sigma}_{\rm stack}^2)^2}{N} [10^{4\,\ln(10)\,s^2}-1] \label{eq:var_stack}.
\end{eqnarray}

Thus we see that the velocity dispersion measured from the stacked spectrum is not given by the
mean log-$\sigma$ value alone, but rather includes a contribution from the width
of the population distribution as well.  In principle, if a good estimator
of $\mathrm{Var}(\sigma_{\mathrm{stack}}^2)$ can be obtained, the system can be closed and solved for
$m$ and $s$ independently.  Although we
attempted to estimate $\mathrm{Var}(\sigma_{\mathrm{stack}}^2)$
via bootstrap resampling within each bin, we found the treatment of observational errors
and varying signal-to-noise ratio among the spectra to be intractable within this framework.
Rather than working further from measurements of stacked spectra,
we proceed to the hierarchical Bayesian method described in the following section.

\subsection{Hierarchical Bayesian Approach}

To constrain the distribution parameters $m$ and $s$ within each redshift--magnitude bin,
we consider the following expansion of the likelihood function $\mathcal{L}(m, s | \{\vec{d}\})$ in the bin:
\begin{eqnarray}
\label{eq:likelihood}
   \mathcal{L}(m, s | \{\vec{d}\}) &=& p(\{\vec{d}\} | m, s) \nonumber \\
                                   &=& \prod_i p(\vec{d}_i | m, s) \\
                                   &=& \prod_i \int \! p(\vec{d}_i | \log_{10}{\sigma})
                                       \, p(\log_{10}{\sigma} | m, s) \, 
                                       \mathrm{d}\log_{10}{\sigma} \nonumber
\end{eqnarray} 
Here $\{\vec{d}\}$ is the set of all spectra in the bin, with each element $\vec{d}_i$ representing the spectrum
of the $i^{\mathrm{th}}$ galaxy. The expression $p(\vec{d}_i | \log_{10}{\sigma})$
is related to the ${\chi}^2(\log_{10} {\sigma})$ function by
\begin{equation}
p(\vec{d}_i | \log_{10}{\sigma}) \propto \exp \left[{-\frac{{\chi}^2_{i}(\log_{10}{\sigma})}{2}} \right]~,
\end{equation}
and $p(\log_{10}{\sigma} | m, s)$ is given by Equation~(\ref{eq:pdf}).
Translating into Bayesian terms, we have a posterior
probability for $m$ and $s$ given by
\begin{equation}
 p(m, s | \{\vec{d}\}) \propto p(\{\vec{d}\} | m, s) \, p(m, s)
\end{equation}
with $p(m, s)$ being the prior probability distribution for $m$ and $s$.  For simplicity, we assume a uniform prior
on $m$ and $s$ over a reasonable range.  In actuality, we find that the likelihood is quite
strongly peaked in each bin, so the exact nature and range of the prior are insignificant.


\subsection{Verification}

To verify the correct functioning of our Bayesian approach,
we make use of data from BOSS plate 3851.  Due to a CCD
failure on one of the two BOSS spectrographs
that temporarily suspended normal survey operations, 500 of the 1000
targets on this plate were plugged and observed for a total
integration time of 7 hours (28 exposures of 15 minutes each)
over the course of several nights ending on 2010 April 12,
denoted within the SDSS-III database by the
modified Julian date (MJD) of 55298.
Subsequent to the replacement of the failed CCD, the entire plate was re-plugged
and observed for a more typical BOSS integration time of 1.75 hours total
on MJD 55302.  The set of re-observed targets allows us to compare
BOSS galaxy spectra of standard survey depth with spectra of the same objects at essentially double
the nominal survey SNR\@.  We use these repeat observations to verify that
our method (1) does not have a signal-to-noise ratio dependent bias in the
estimation of velocity-dispersion distribution parameters, and (2)
reproduces the known distribution of velocity dispersions within
a controlled sample, as measured from the high-SNR set of spectra.

\begin{figure}[t]
\centering
\plotone{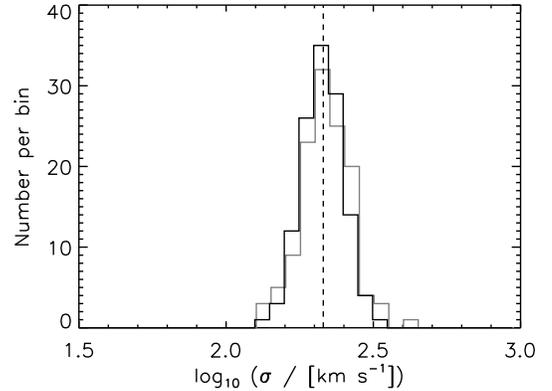}
\caption{\label{fig:vdTestHist}
Engineered sub-sample Gaussian histogram in $\log_{10} \sigma$ constructed
using measurements from 7-hour
BOSS observations \textit{(black)}, with histogram of same
sub-sample using velocity dispersions measured from 1.75-hour
observations \textit{(gray)}.  The two histograms
have been given a slight relative horizontal offset, for display
purposes.  The vertical dashed line
indicates the mean $\log_{10} \sigma$ value of 2.33 for the sub-sample.
Note the relative broadening of the 1.75-hour histogram due to the effects
of observational error.}
\end{figure}

Between the deep and shallow re-observations, there are 308 galaxies
which have equal redshifts (within $\Delta z = \pm 0.005$)
and classifications for both observation dates.
Since the sample is heterogeneous in magnitude and redshift,
we select a sub-sample with a controlled distribution in velocity dispersion.
We restrict our attention to galaxies that have their individual velocity
dispersions measured at SNR of 10 or more from the 7-hour observations,
and that have redshifts between 0.4 and 0.6.  We then select a random
sub-sample of 125 galaxies from this set so as to have a Gaussian histogram in
$\log_{10} \sigma$ with a mean of $m = 2.33$ and an intrinsic
RMS scatter of $s = 0.07$.  The histogram of this sub-sample, along
with the histogram of the same sample as constructed
from galaxy-by-galaxy measurements using
the 1.75-hour observations, is shown in Figure~\ref{fig:vdTestHist}.

The frequentist formulas given by Equations (\ref{eq:sigma_stack}) and (\ref{eq:var_stack})
do not account for observational error, and hence we do not use them
to solve for $m$ and $s$ estimates for our relatively low signal-to-noise BOSS
survey data.  However, our subsample
of high signal-to-noise 7-hour observations
allows us to test them, which we do before proceeding to the
verification of our Bayesian analysis framework.
First, we use Equation~(\ref{eq:sigma_stack}) with a mean of 2.33 and an intrinsic scatter of 0.07 
to predict a value of $\sigma_{\rm stack}=\rm 219\,km\,s^{-1}$,
which is in very good agreement with the result of $\rm (222 \pm 12) \,km\,s^{-1}$
that we obtain by 
fitting the stacked spectrum of this set of 125 galaxies directly. Similarly,
we predict $[\mathrm{Var}(\sigma_{\mathrm{stack}}^2)]^{1/4} = 40 \rm\,km\,s^{-1}$ from 
Equation~(\ref{eq:var_stack}), which is in reasonable agreement
with the value of $46 \rm \,km\,s^{-1}$ obtained through a
bootstrap resampling process.  In both cases the agreement is not exact
because there is still some observational error even in the
7-hour data, but as mentioned above, we will pass to the
Bayesian framework to quantify these effects.

\begin{figure}[t]
\centering
\plotone{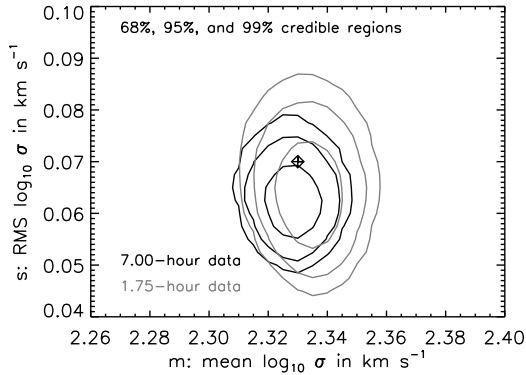}
\caption{\label{fig:bayesVerify}
Credible-region contours of constant posterior probability density
for $m$ and $s$ parameters measured from the engineered test sub-sample
of galaxies observed with both 7-hour integrations \textit{(black)}
and 1.75-hour integrations \textit{(gray)}.  The symbol is the location
of the parameters chosen for the construction of the test sub-sample.
The offset in $s$ between the contours and the symbol
is a result of the proper deconvolution of
observational uncertainty that is implemented by the
Bayesian method.
}
\end{figure}

We next carry out the estimation of the $m$ and $s$ parameters
of the selected sub-sample of objects,
using the Bayesian approach described above, for both
the 7-hour and the 1.75-hour data sets.  Figure~\ref{fig:bayesVerify}
shows the resulting posterior probability density for these parameters
as estimated from both data sets.
As expected, we
see that the posterior PDF is tighter for the 7-hour data.
More importantly, we see no significant bias in the posterior PDF between the
low-SNR and high-SNR data sets.  This is especially significant
for the estimation of the $s$ parameter: if we were not handling
our observational uncertainties correctly, we might expect to infer
a broader intrinsic distribution (higher $s$ value) from the noisier
data, but this not the case.  We also see that the
parameters used to engineer the subsample are recovered
with no significant bias in $m$.  We see a slight offset
of the 7-hour maximum posterior $s$ value from the input value
used to engineer the sample.  This is in the direction and of the
size to expected given the
observational error of the 7-hour individual-spectrum
velocity dispersion measurements, which have an RMS signal-to-noise
of about 17.  This corresponds to an observational broadening of
about 0.025\,dex in the engineered histogram of Figure~\ref{fig:vdTestHist},
which is deconvolved
by the Bayesian parameter estimation procedure to give the lower
recovered $s$ value seen in Figure~\ref{fig:bayesVerify}.

\begin{figure}[thbp]
\centering
  \plotone{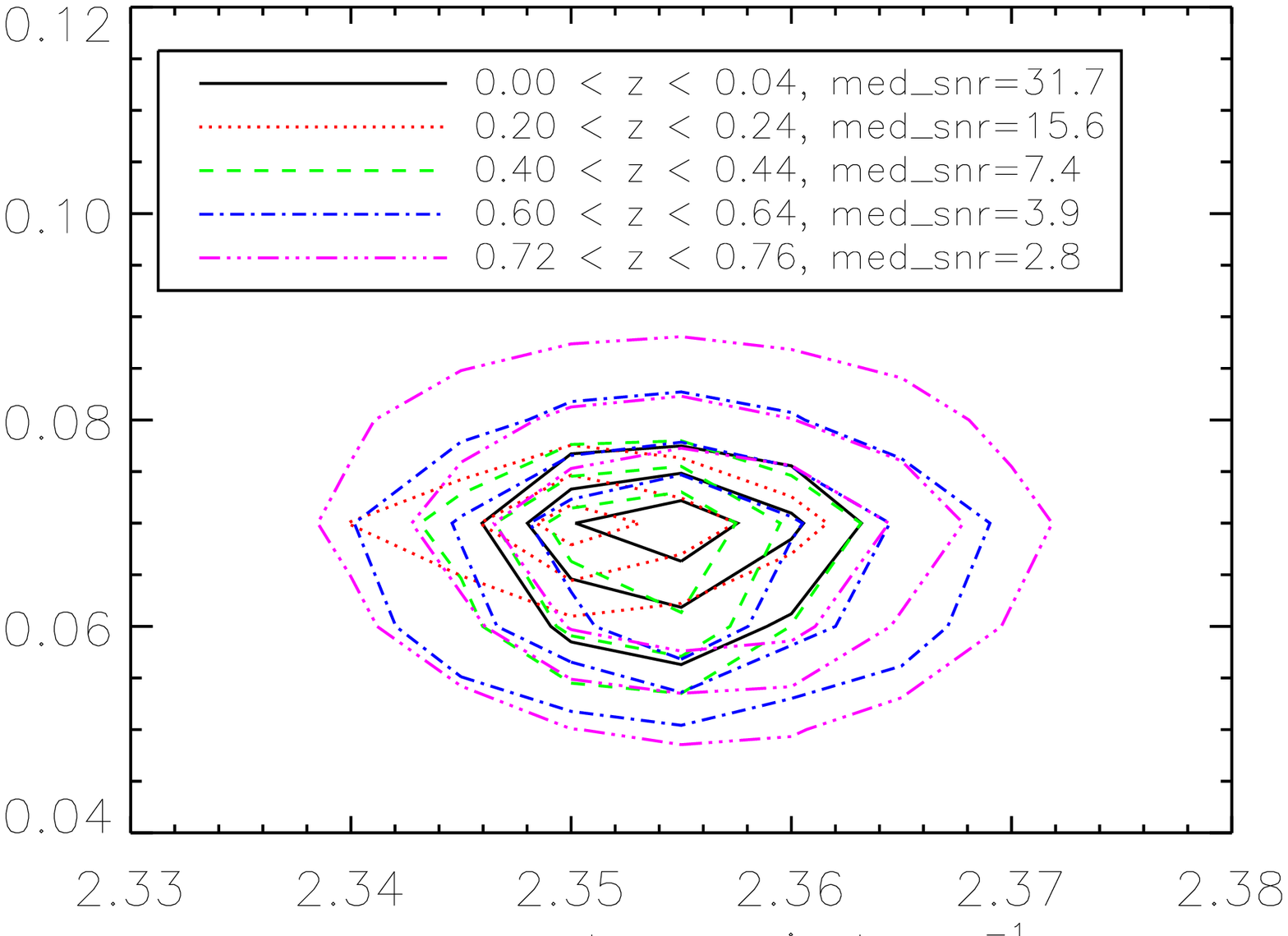}
  \vspace{4 mm}
  \caption{\label{fig:vdisp_verify1}
    Contours of constant posterior probability density (68\%, 95\%, and 99\%) for m and s parameters 
    obtained from a controlled sub-sample of 152 galaxies in 5 different redshift bins 
    with gradually reduced SNRs.}
\end{figure}

\begin{figure}[hbtp]
\centering
  \plotone{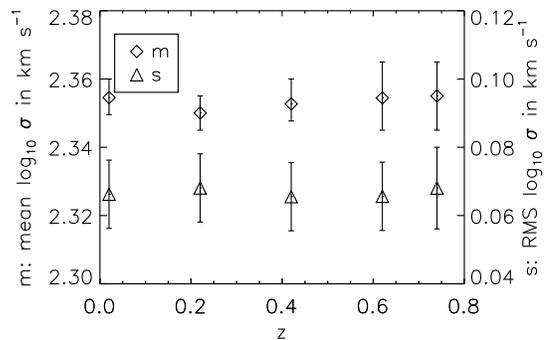}
  \caption{\label{fig:vdisp_verify2}
    The best estimated m (Diamond) and s (Triangle) values for the controlled sub-sample of 152 galaxies 
    at 5 different redshift bins.}
\end{figure}

Another concern is that there might be a systematic bias with redshift, 
since the spectral regions used by \texttt{vdispfit} in fitting for
velocity dispersions (rest frame wavelength range from 4,100\AA\ to 6,800\AA)
move to the redder and noisier parts of the spectrum as the redshift gets higher. 
In order to test this, we construct another controlled sub-sample with 152 galaxies 
of redshift $z < 0.2$ and very high SNR.  
Then we take the best-fit template combination models of those 152 galaxies returned by \texttt{vdispfit} and 
redshift them to progressively higher redshift bins, giving them a uniform random distribution
over a bin width of $\Delta z = 0.04$ in each case (to match our actual binning). 
At each new redshift, the model spectra are added to sky-subtracted BOSS sky fibers
to simulate realistic survey noise, and scaled individually in flux to give a typical median
SNR at that redshift bin. We then analyse the simulated
redshifted samples with our Bayesian method to estimate the posterior PDF
of $m$ and $s$. The results are shown in Figure~\ref{fig:vdisp_verify1} and Figure~\ref{fig:vdisp_verify2}, for 5
separate redshift bins. We see that the recovered parameters are consistent
within observational error across all redshifts, with no apparent redshift-dependent bias.

Finally, to rule out any significant dependence of our
measurement on airmass and fiber position within
the BOSS spectrographs, we make use of data from plates 3615, 3647,
4238 and 4239.  These four plates cover roughly the same area of sky,
but with different plate drillings that place the same objects in very different
fibers within the spectrograph system.  They were also observed over a range
of different airmasses on multiple nights.  From these plates, we construct several sub-samples
of spectra, all of which include the same galaxies, but are drawn from different
plates and/or observations.  As with the previous tests, we recover consistent
estimates of $m$ and $s$ from the analysis of all these samples.

Based on the above three tests, we conclude that our method recovers
accurate estimates of the population velocity-dispersion distribution parameters.

\subsection{Magnitude Error Correction}

Our method of determining $p(\vec{d}_i | \log_{10} \sigma)$ incorporates an
explicit marginalization over redshift error, and propagates
all observational uncertainty in the velocity dispersion of
a given galaxy.  Our binning in redshift and absolute magnitude
introduces additional error possibilities that we must
account for.  In the case of redshift, the errors are negligible
relative to the bin width of $\Delta z = 0.04$,
and are unlikely to contribute any
artificial broadening to our determination of the redshift
dependence of $m$ and $s$.  The absolute magnitude errors are, 
however, non-negligible in comparison to the magnitude bin width
of $\Delta \rm M_V = 0.1$, and thus we use the following technique to estimate
and compensate for the broadening effect of the
observational scattering of galaxies between
absolute-magnitude bins (see Figure~\ref{fig:gal_dist}).

Suppose $(m, s)$ are the true values within a bin, and $(m_1, s_1)$ are the values
that we determine in the presence of absolute-magnitude errors.
We assume that
\begin{eqnarray}
  m_1&=&m + {\delta}m  \\
  s_1^2&=&s^2 + {\delta}s^2~,
\end{eqnarray}
where ${\delta}m$ and ${\delta}s$ are the biases introduced by magnitude errors.
To estimate and remove these biases,
we add additional random errors to all our galaxy absolute magnitudes $M_V$ to give
\begin{equation}
  M_V^\prime = M_V + {\epsilon} {\delta}M_V ~,
\end{equation}
where ${\epsilon}$ is a normally distributed random number with mean 0 and standard deviation 1,
and ${\delta}M_V$ are the galaxy-by-galaxy absolute-magnitude
errors estimated by \texttt{sdss2bessell} (propagated from SDSS \textit{ugriz}
apparent magnitude errors).  We repeat our analysis, binning instead in $M_V^\prime$,
and denoting the new distribution parameter results by $m_2$ and $s_2$.
We assume these new determinations are related to $m_1$ and $s_1$
in the same way as $m_1$ and $s_1$ are related to $m$ and $s$,
which implies that
\begin{align}
  m_2&=m + 2\,{\delta}m  \\
  s_2^2&=s^2 + 2\,{\delta}s^2
\end{align}
Thus the biases due to absolute magnitude errors ${\delta}m$
and ${\delta}s$ can be removed to yield
\begin{eqnarray}
  m&=&2\,m_1-m_2 \\
  s&=&\sqrt{2\,s_1^2-s_2^2}
\end{eqnarray}
In practice, we find typical values for ${\delta}m$ of 0.01, and for ${\delta}s$ of 0.04. 

\section{Results: Evolution of the Velocity-Dispersion Function}
\label{sec:evol}

In this section we present the results of the application of our
hierarchical Bayesian velocity-dispersion distribution measurement
technique to the approximately 103,000 galaxies in our LOZ sample and
330,000 galaxies in our CMASS sample.

\subsection{LOZ Sample}

\begin{figure*}[t]
\centering
   \plottwo{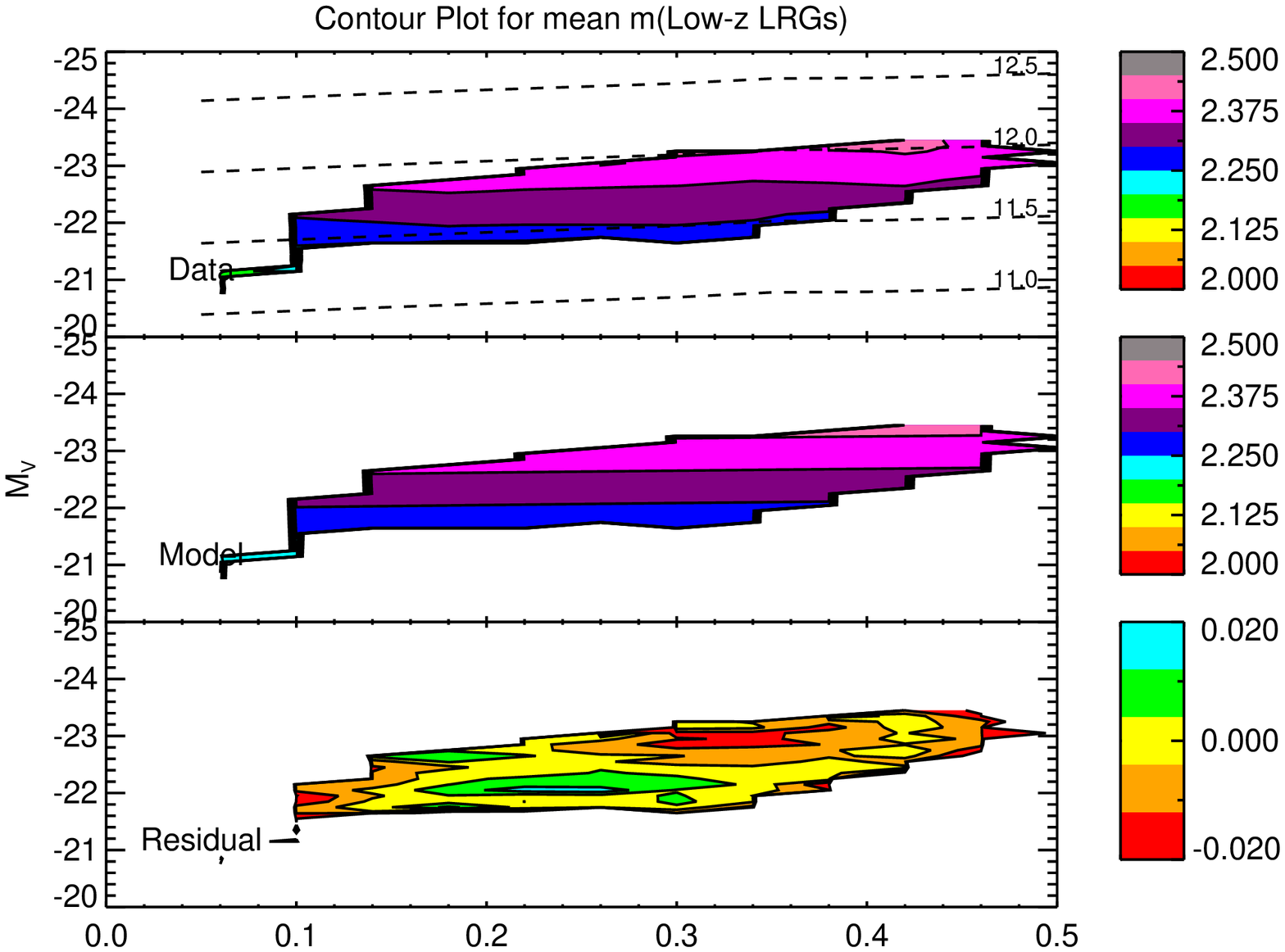}{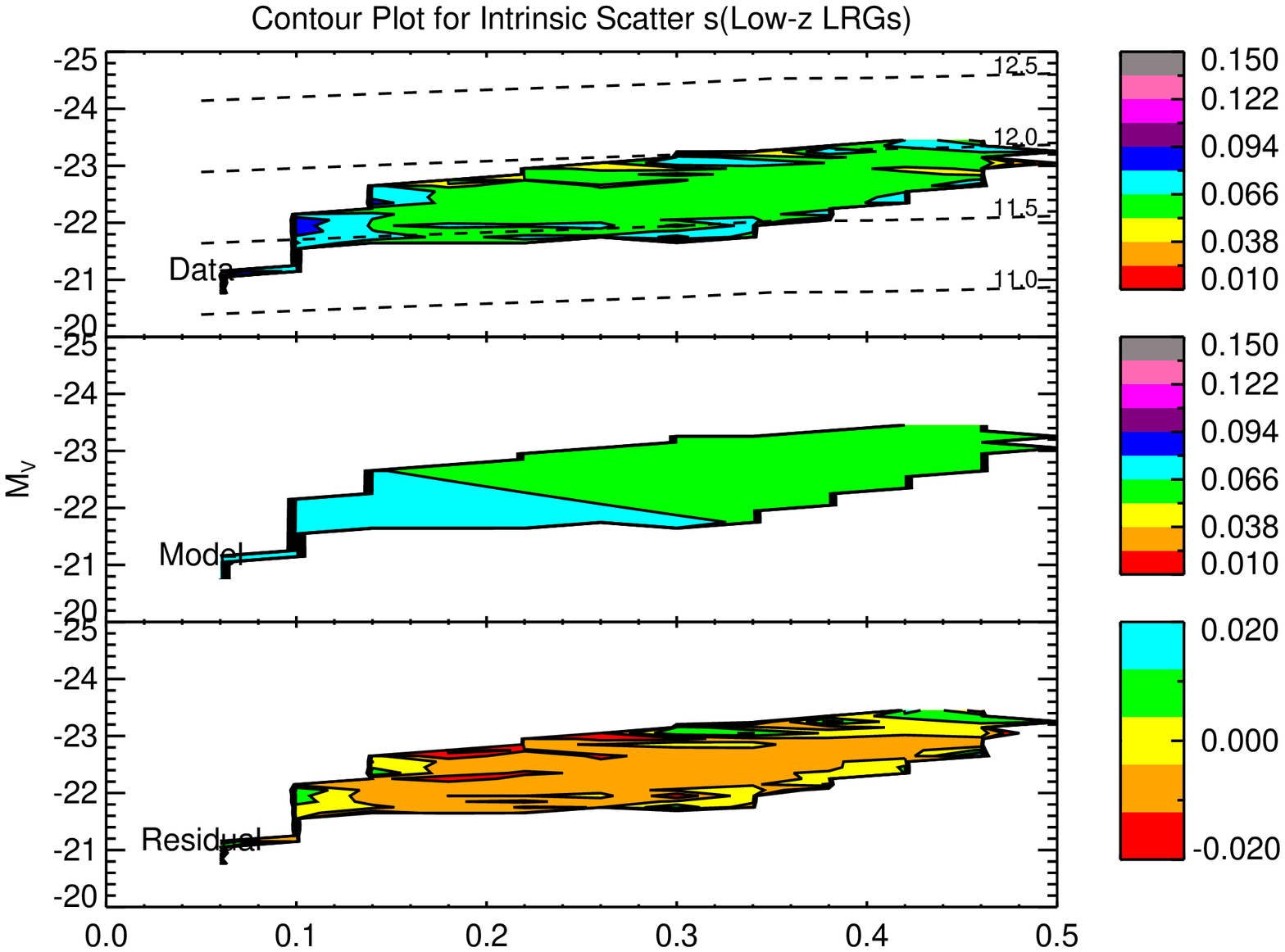}
\caption{\label{fig:loz}
Contour plots of $m$ \& $s$ for LOZ sample galaxies.  Top panels show the map of maximum posterior probability
over the range of the plane with bins containing at least 100 galaxies. Middle panels show low-order bivariate model fits
to these maps constructed as described in the text, and residuals (top minus middle) are shown in the bottom panels.
Dashed lines in the top panels show tracks of constant stellar mass from the LRG population
model of \citet{Maraston09}.
}
\end{figure*}

\begin{figure*}[t]
\centering
   \plottwo{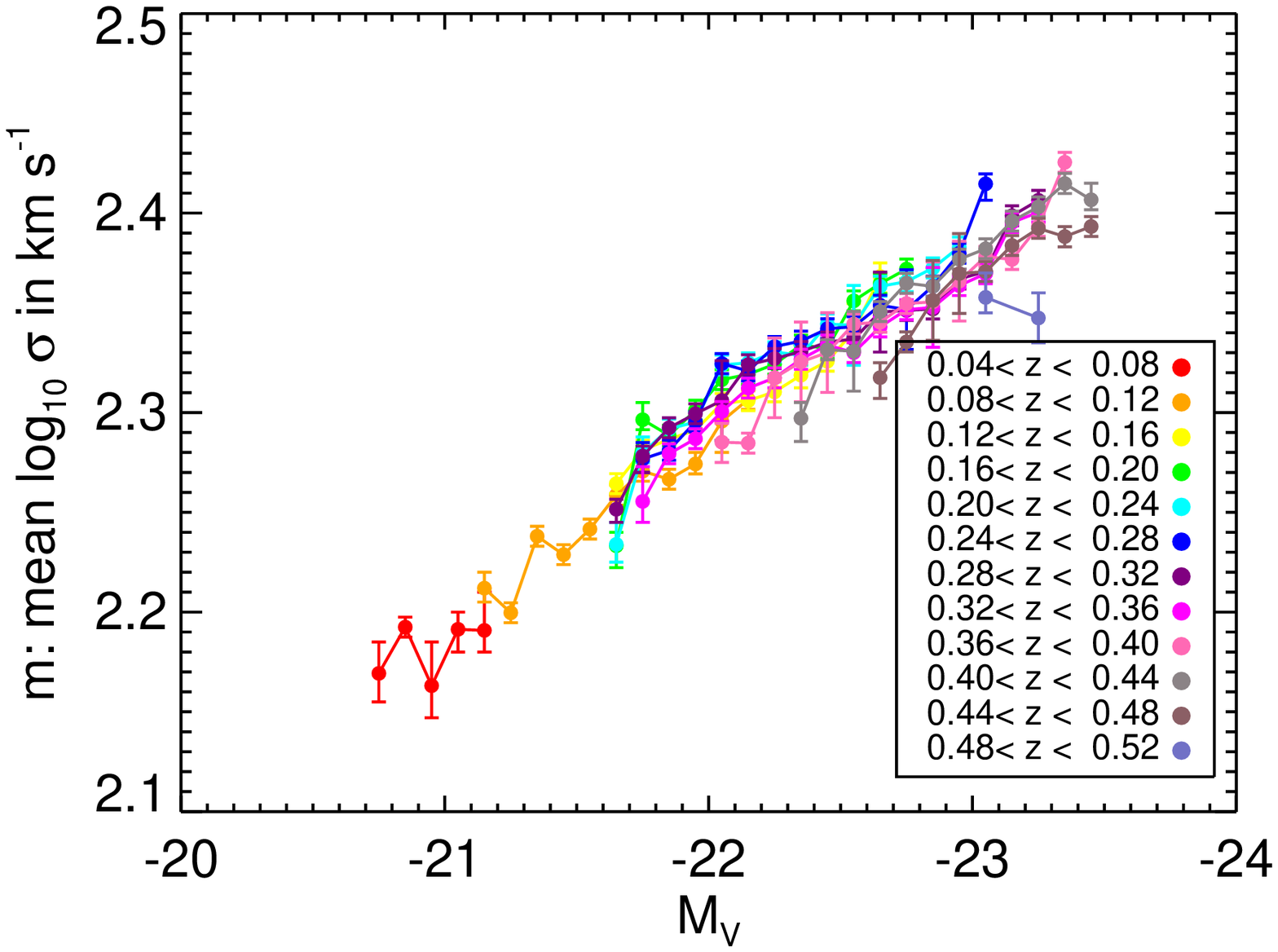}{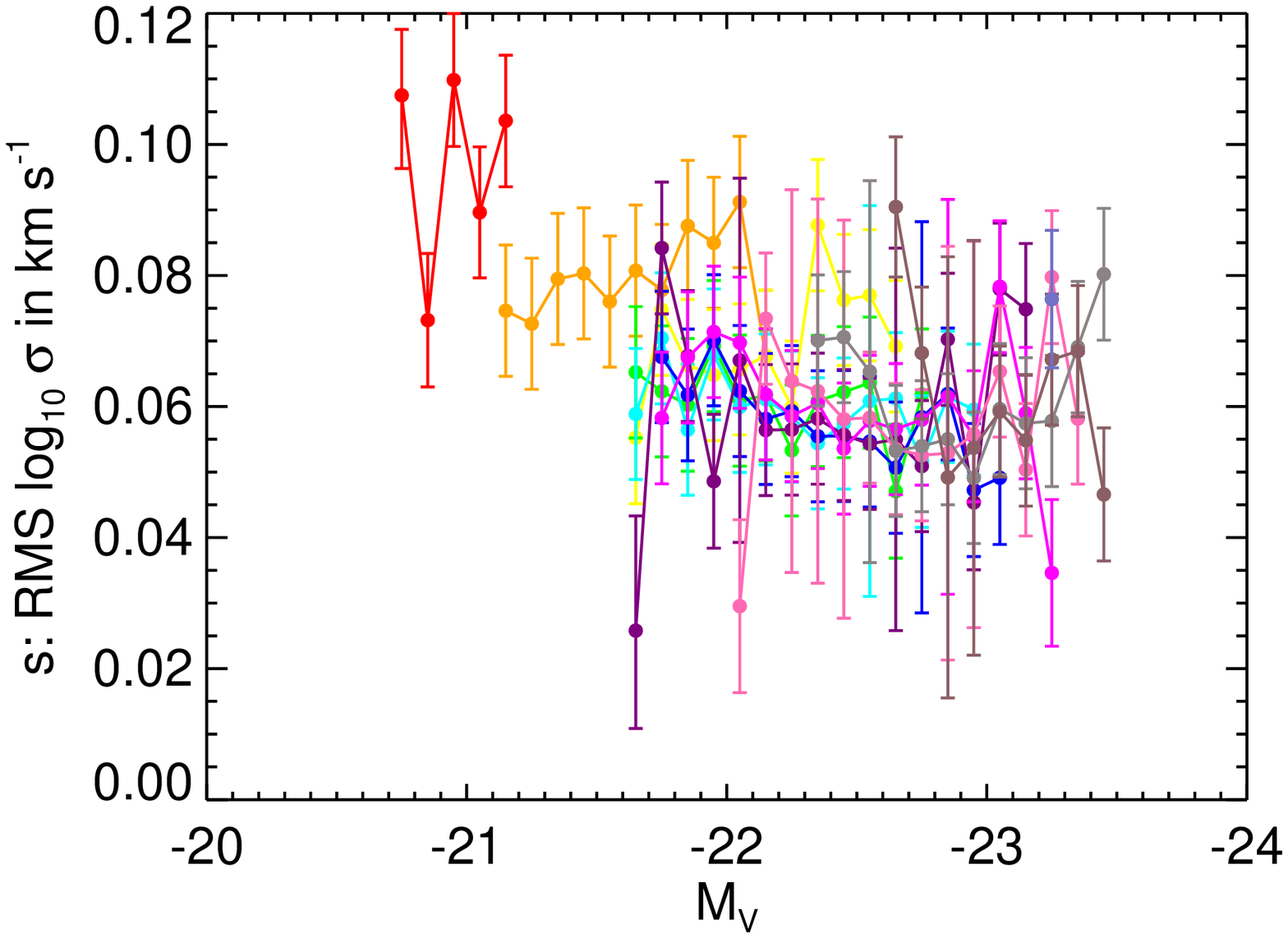}
   \plottwo{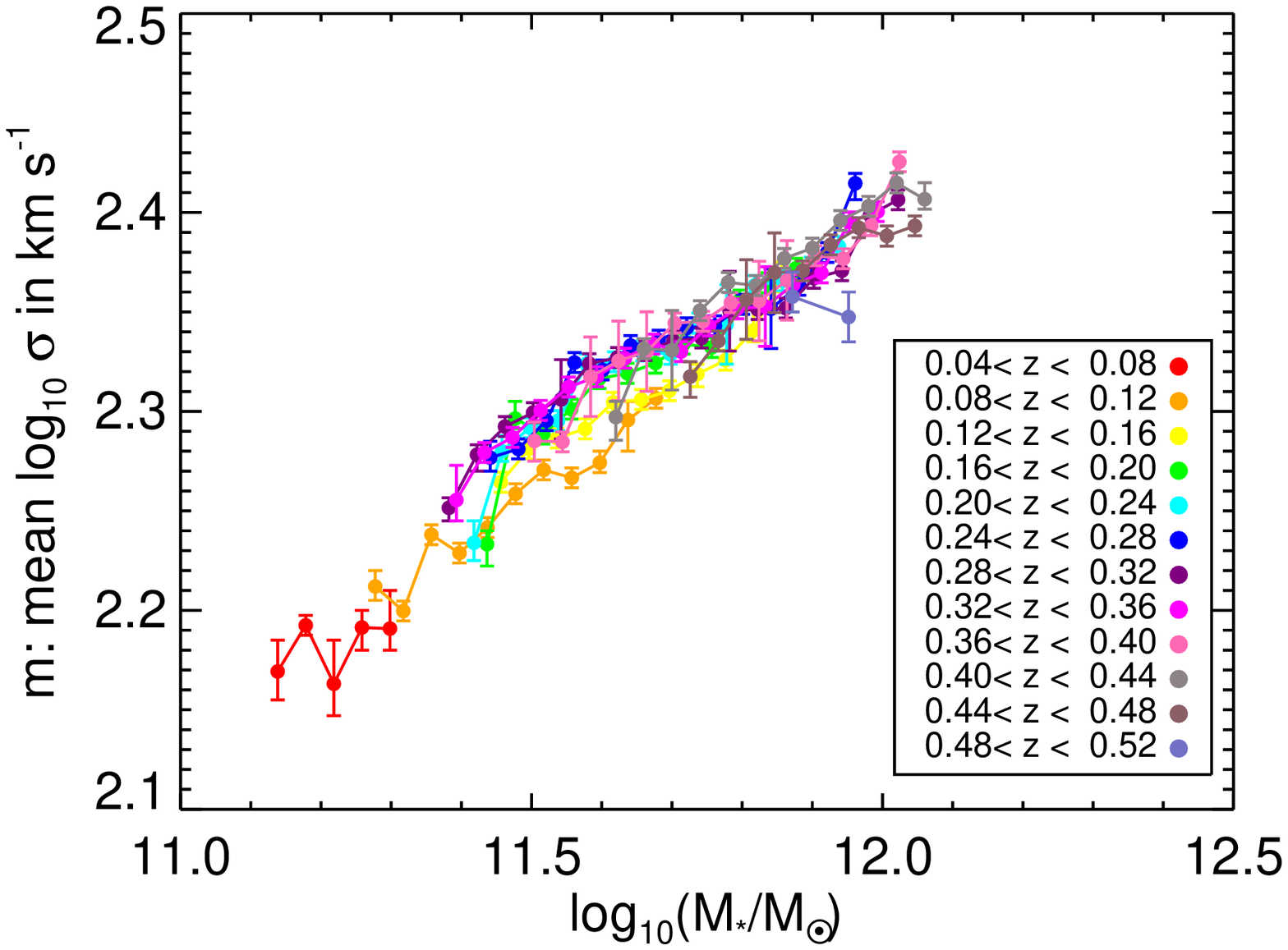}{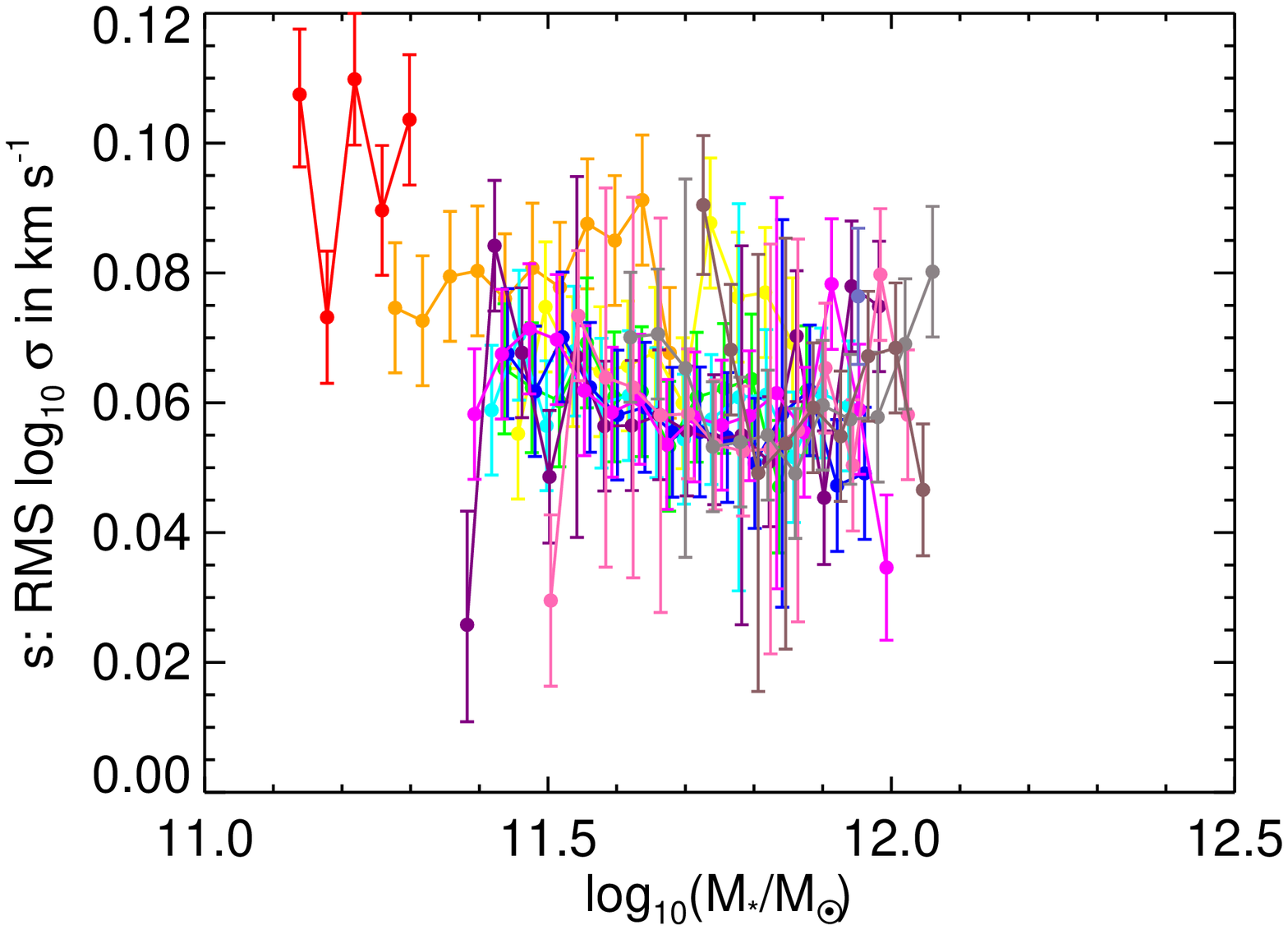}
\caption{\label{fig:scatter-loz}
Scatter plots of $m$ \& $s$ versus $M_V$ (top two panels) and $\log_{10}(M_*/M_\sun)$ (bottom two panels) 
for LOZ sample galaxies in different redshift ranges.
}
\end{figure*}

\begin{figure*}[t]
\centering
	\plottwo{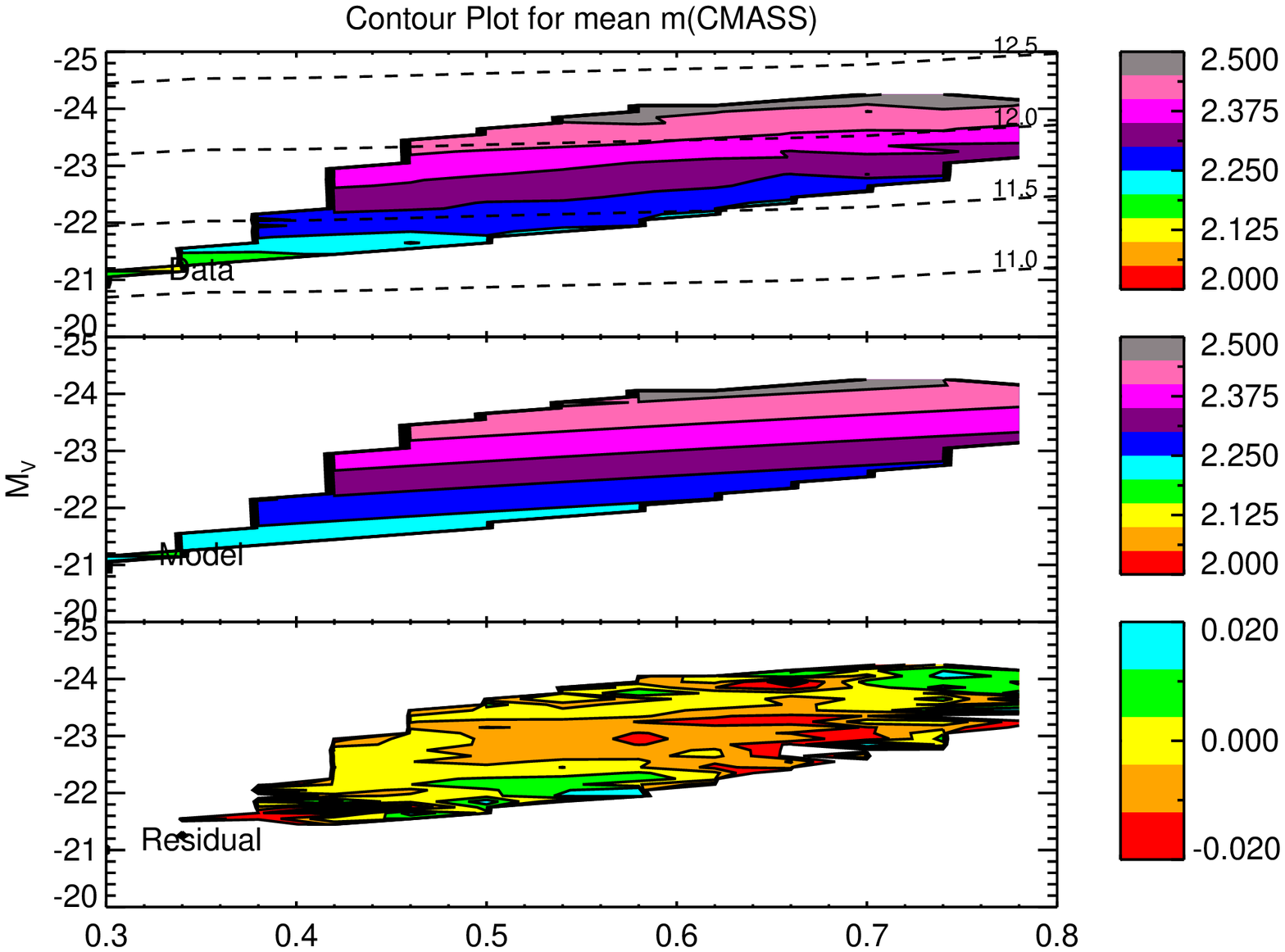}{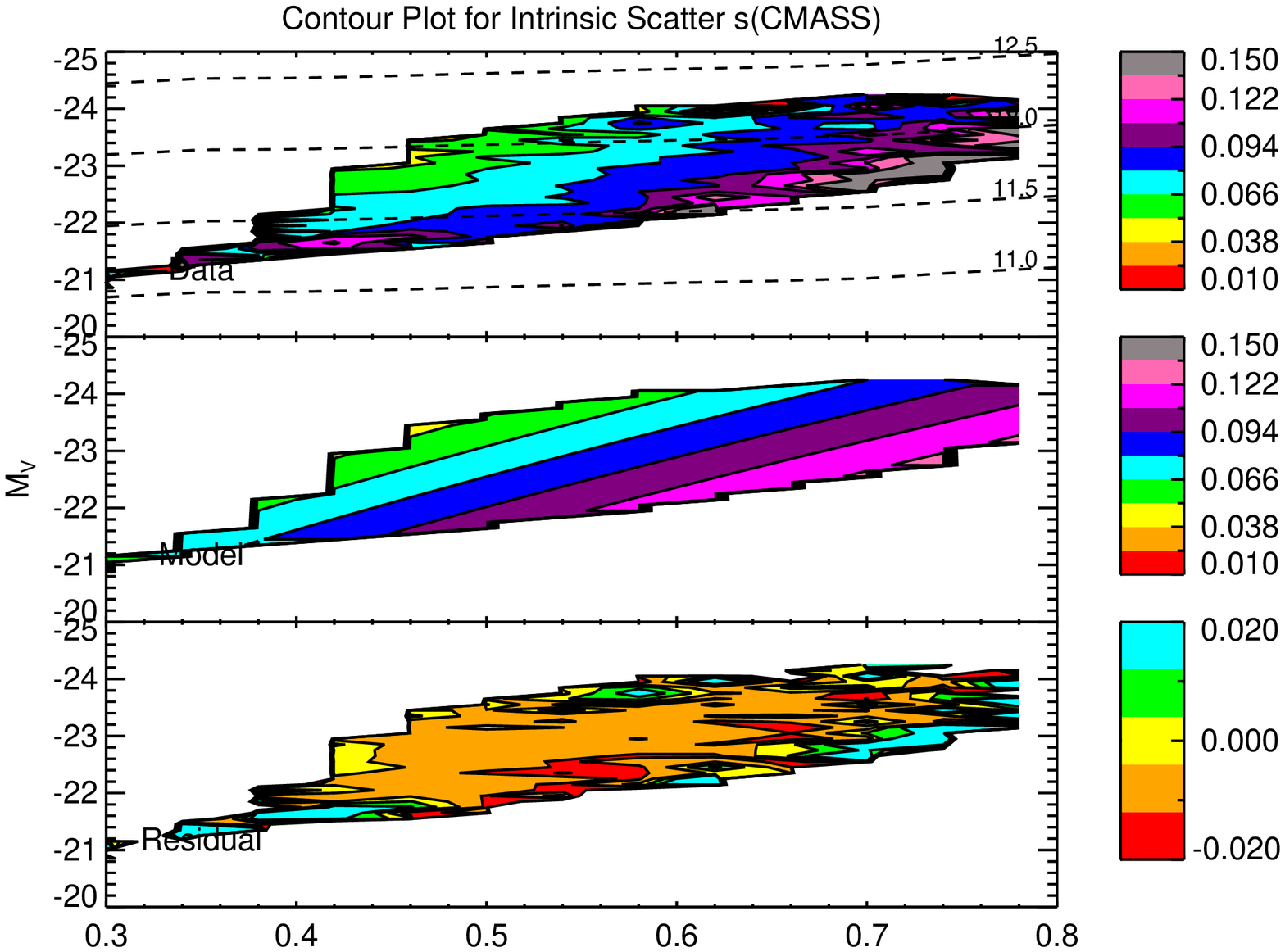}
\caption{\label{fig:CMASS}
The same as Figure~\ref{fig:loz} but for CMASS galaxies.
}
\end{figure*}

\begin{figure*}[t]
\centering
   \plottwo{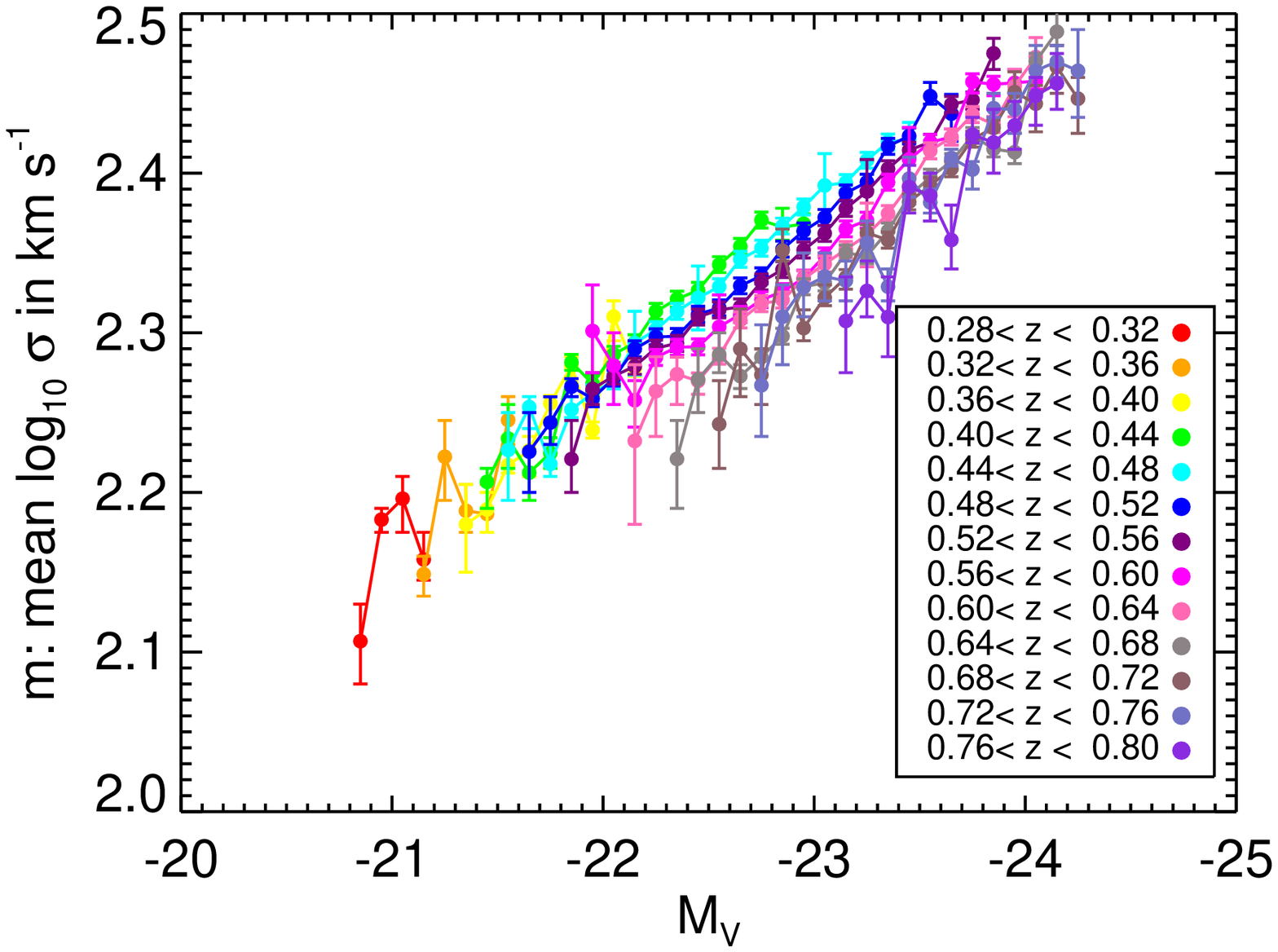}{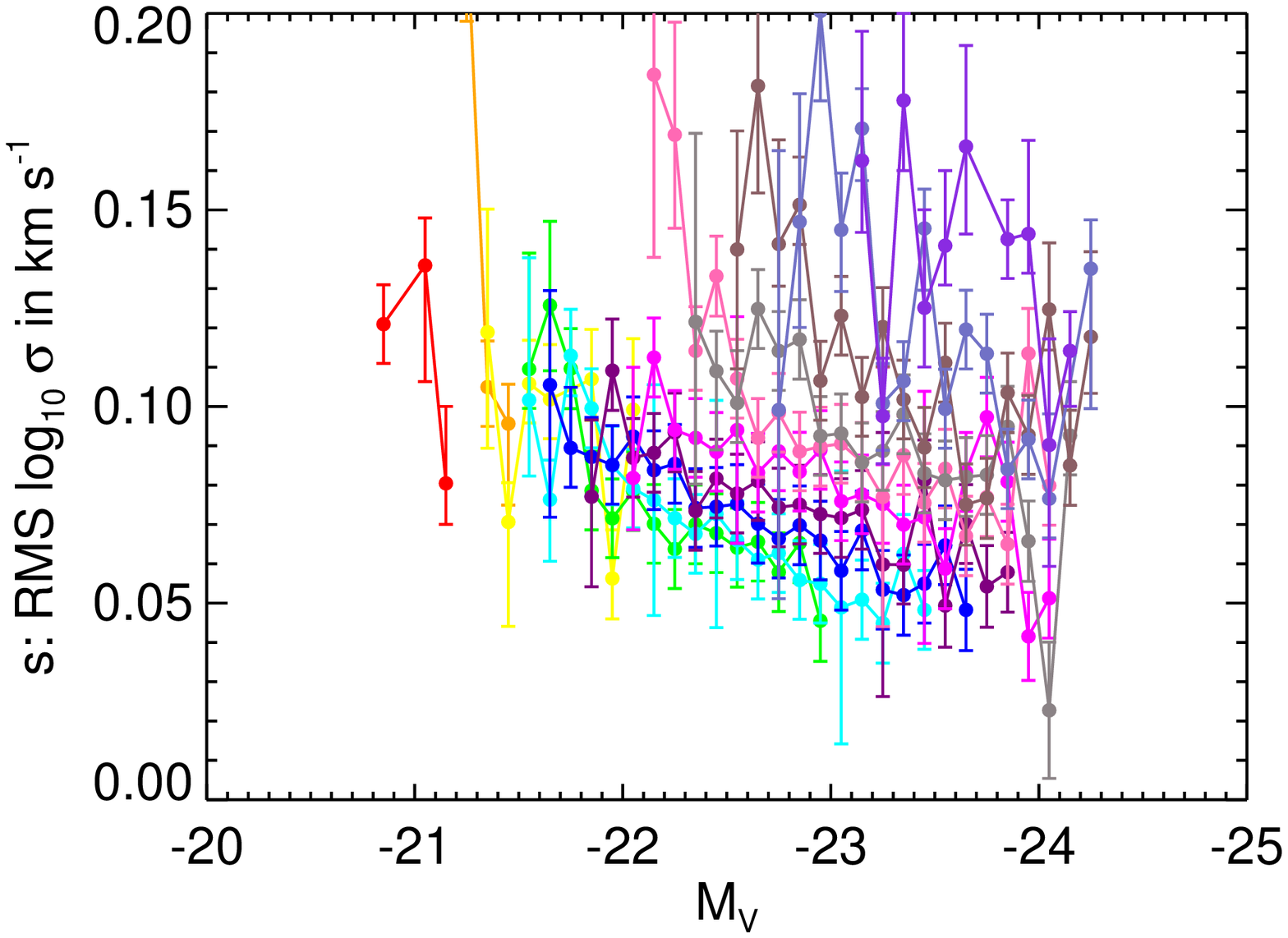}
   \plottwo{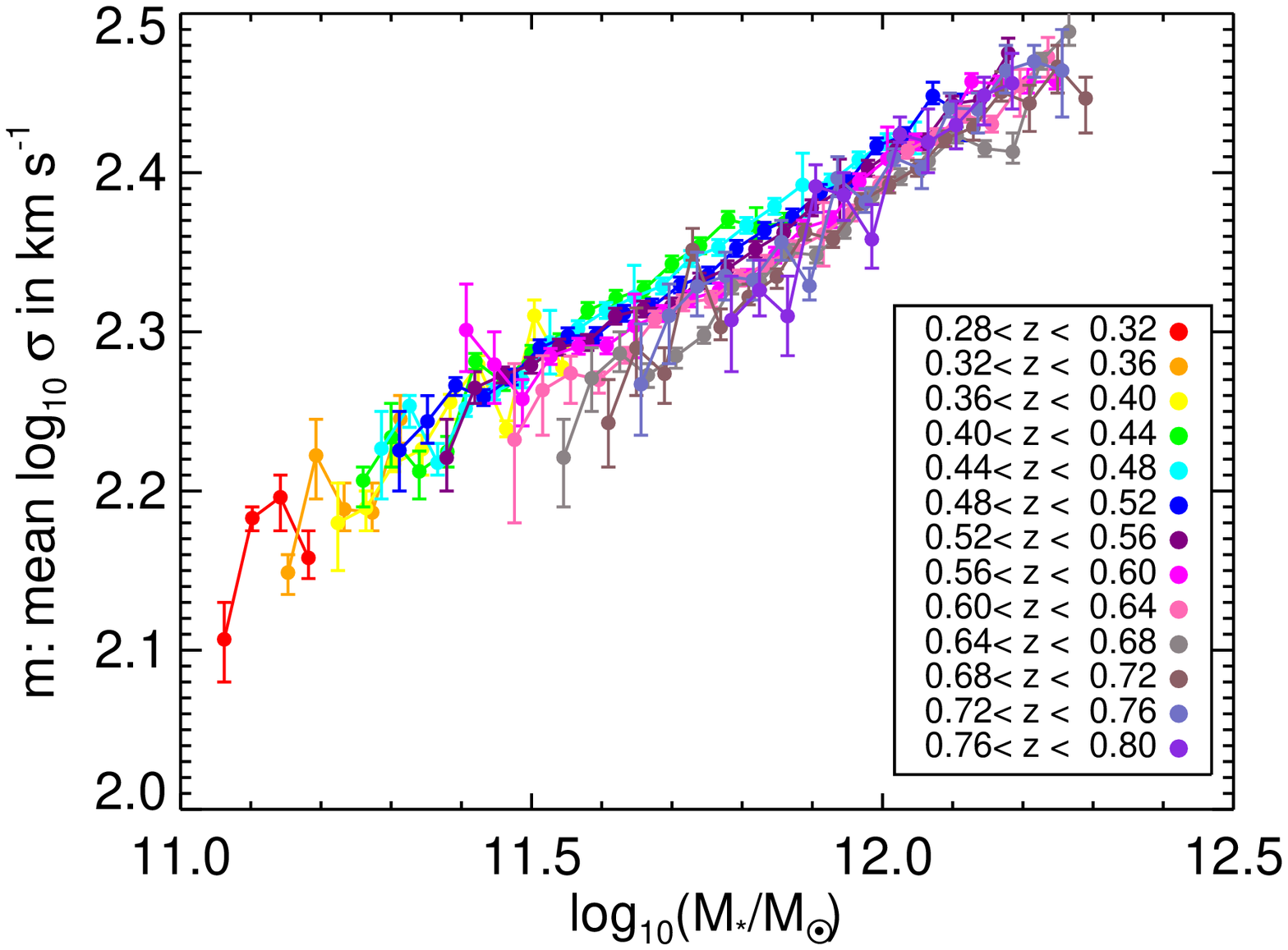}{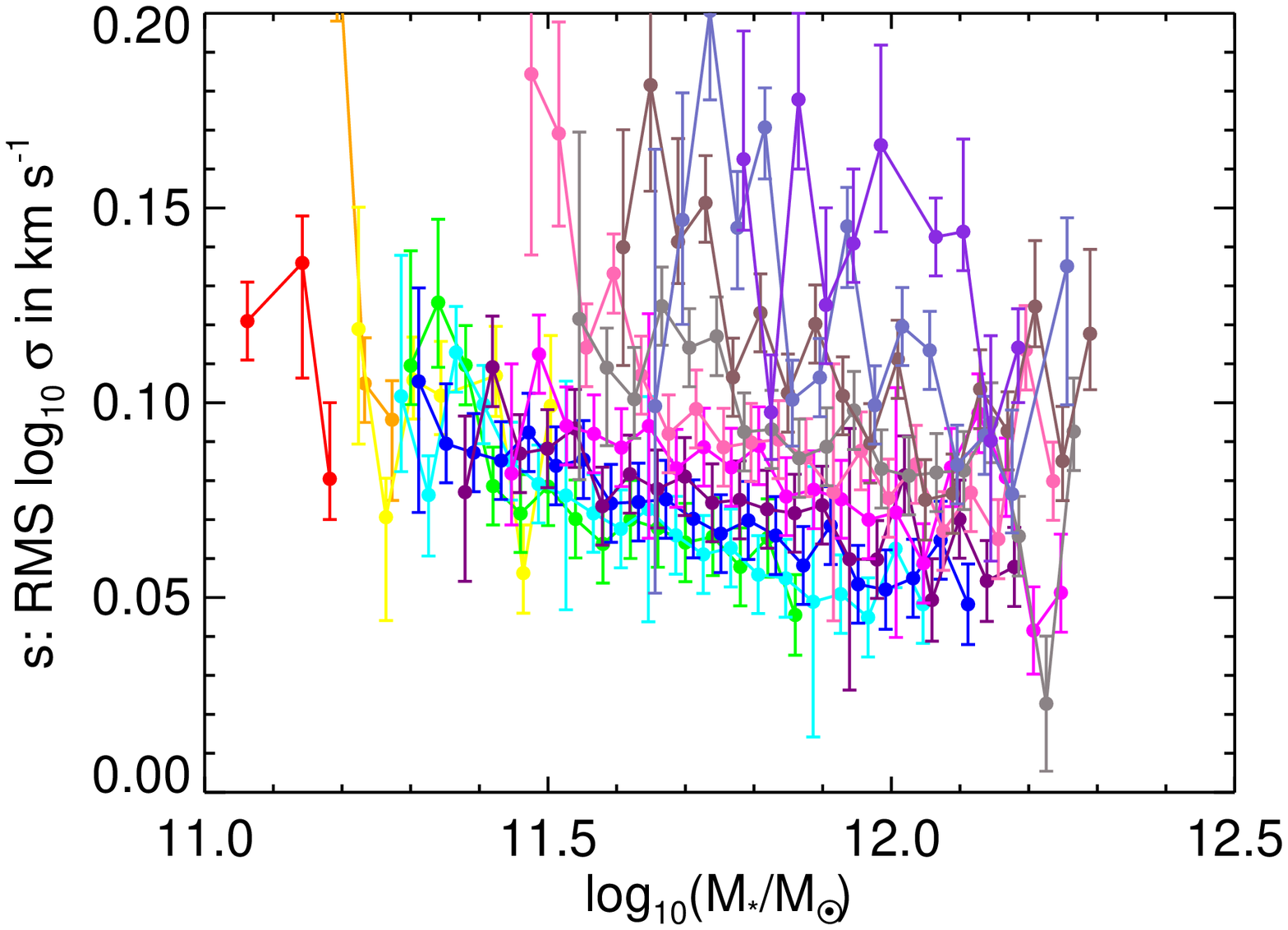}
\caption{\label{fig:scatter-CMASS}
The same as Figure~\ref{fig:scatter-loz} but for CMASS galaxies. 
(Note that the scales of these panels are expanded relative to Figure~\ref{fig:scatter-loz}.)
The increase of intrinsic scatter with redshift can be seen in the right-hand figure.}
\end{figure*}

The LOZ sample extends to $z \approx 0.5$. The 2D contour plots
of $m$ and $s$ (Figure~\ref{fig:loz}) and scatter plots in different redshift bins (Figure~\ref{fig:scatter-loz})
show that the mean $m$ is strongly correlated with absolute magnitude, while 
the intrinsic scatter $s$ shows no significant variation.
Tracks of constant stellar mass assuming the LRG stellar population model of \citet{Maraston09}
have also been over-plotted in Figure~\ref{fig:loz}, and used to convert
from an absolute-magnitude to a stellar-mass baseline in Figure~\ref{fig:scatter-loz}.
Galaxies in the LOZ sample have estimated stellar masses between approximately $10^{11} M_\sun$ and $10^{12} M_\sun$. 

To quantify the variation of the $m$ and $s$ parameters with redshift and absolute
magnitude, we consider a simple model specified by:
\begin{eqnarray}
	m^0 =& \rm A_m^0 M_V + B_m^0 \log_{10}(1+z) &+ C_m^0  \\
	s^0 =& \rm A_s^0 M_V + B_s^0 \log_{10}(1+z) &+ C_s^0  ~,
\end{eqnarray}
with the ``0'' superscript denoting the LOZ sample specifically.
Performing a linear least squares fit to the individual bin data points,
we obtain
\begin{align}
	\nonumber A_m^0&=-0.0880\pm0.0012 & A_s^0&=0.006\pm0.002 \\
	\nonumber B_m^0&=-0.087\pm0.018 & B_s^0&=-0.08\pm0.03 \\
	C_m^0&=0.37\pm0.02 & C_s^0&=0.20\pm0.04
\end{align}
We can translate the resulting scaling into the standard
form for the FJR, with luminosity $L \propto \sigma^x$ by
recognizing that $x = -0.4 / A_m^0$.  The resulting value
of $x = 4.55 \pm 0.06$ is in reasonable agreement with
the canonical local-universe value of $x = 4$.
Thus, BOSS low-z LRGs define an FJR whose slope and
scatter has little dependence on redshift and luminosity;
there is correspondingly little evidence for dynamical
evolution in this sample since roughly $\rm z=0.5$.

\subsection{CMASS Sample}

The CMASS galaxy sample extends from $z \approx 0.3$ to $z \approx 0.8$.
The results of our $\sigma$ distribution parameter measurements
are shown in Figures \ref{fig:CMASS} and~\ref{fig:scatter-CMASS},
once again using tracks of constant stellar mass
based on the \citet{Maraston09} population model.
Using the same model form as used for the LOZ sample above,
\begin{eqnarray}
	m^1 =& \rm A_m^1 M_V + B_m^1 \log_{10}(1+z) &+ C_m^1 \\
	s^1 =& \rm A_s^1 M_V + B_s^1 \log_{10}(1+z) &+ C_s^1
\end{eqnarray}
(with the ``1'' superscript denoting the CMASS sample specifically),
and again doing a linear least-squares fit, we find that 
\begin{align}
	\nonumber A_m^1&=-0.1128\pm0.0010 & A_s^1&=0.0263\pm0.0016 \\
	\nonumber B_m^1&=-0.77\pm0.02 & B_s^1&=0.82\pm0.04 \\
	\nonumber C_m^1&=-0.089\pm0.019 & C_s^1&=0.52\pm0.03
\end{align}

In the case of the CMASS sample, 
the FJR is still apparent, but the scaling
exponent in $L \propto \sigma^x$ is now $x = 3.55 \pm 0.03$.
This observation that the FJR becomes ``shallower'' at higher redshift
can be interpreted in terms of mass-dependent star-formation history 
\citep[e.g.,][]{cowie_96, Alighieri05},
with less massive (lower $\sigma$) galaxies having undergone
more recent star formation and thus fading more rapidly with cosmic time
relative to more massive galaxies.

There is a clear evolution in the zero-point of the $m$ versus $M_V$ relation 
(upper left panel in Figure~\ref{fig:scatter-CMASS}) with redshift. 
This evolution is essentially eliminated in the lower left panel
of Figure~\ref{fig:scatter-CMASS}, which translates to a baseline
of constant stellar mass.
Hence the evolution of the $m$ versus $M_V$ relation in the
CMASS sample is consistent with passive stellar evolution. 

\begin{figure}
\centering
\plotone{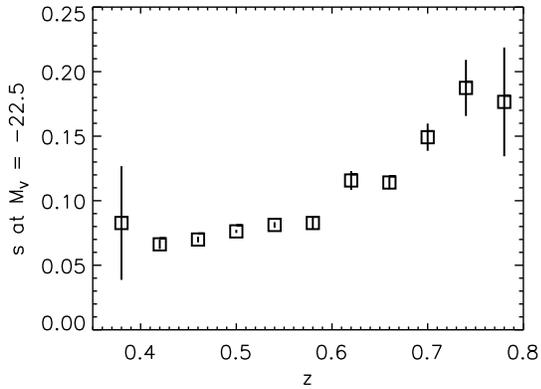}
\caption{\label{fig:s_zeropt}
Variation of the intrinsic width $s$ of the
CMASS population distribution in $\log_{10} \sigma$
as a function of redshift.}
\end{figure}

It can easily be seen from Figures \ref{fig:CMASS} and~\ref{fig:scatter-CMASS}
that $s$ is no longer constant with redshift at fixed luminosity or stellar mass.
The significance of this result is encapsulated in the
non-zero value of $B_s^1 = 0.82 \pm 0.04$ given above.
To quantify this result in more detail, we fit the $s$ versus $M_V$
relation with a linear model at each redshift bin, and plot the
zero-point of this relation as a function of redshift
in Figure~\ref{fig:s_zeropt}.
We see that within the CMASS sample, the intrinsic width
$s$ of the velocity-dispersion function at fixed magnitude
or stellar mass \textit{decreases} with cosmic time (i.e.,
broader distribution width at higher redshift),
especially at redshifts $z > 0.6$.
This is consistent with our tentative detection of evolution
in the FJR slope between the LOZ and CMASS samples,
in the sense that a given range in luminosity
encompasses a larger range of velocity dispersions
at higher redshift, but the signal is too large
to be explained by this effect alone
(since the FJR slope is not seen to evolve
significantly \textit{within} the CMASS sample alone).
We are therefore seeing increased dynamical
heterogeneity at fixed luminosity in the CMASS
sample at higher redshifts.

We note that the apparent increase in the intrinsic $\sigma$ distributions
at high redshift cannot be explained in terms of surface-brightness selection effects.
Through the FP or Kormendy relations, velocity dispersion at fixed
luminosity is correlated with surface brightness.  At the high-redshift
end of the CMASS sample, we can expect a degree of incompleteness at
both ends of the surface-brightness distribution.  On the one hand,
relatively low surface-brightness
galaxies will have fainter magnitudes within the BOSS spectroscopic fiber,
and will thus be less likely to be targeted, and less likely to have
confident and correct spectroscopic redshift measurements even if targeted.
On the other hand, relatively high surface-brightness galaxies
(again, at fixed luminosity) run the risk of being unresolved in
star--galaxy separation.  Consequently, we might expect the
distribution of velocity dispersion at fixed magnitude to
be made more narrow at high redshift by these considerations,
which goes in the opposite sense to the trend we detect.

\subsection{Application to Individual Spectra}
\label{sec:prior}

Our results characterize the dynamical properties of the population of LRGs
targeted by BOSS\@.  The parameters of the population can in turn
be used to inform our estimates of the velocity dispersion values
of individual noisy BOSS spectra.  For this application, we want
to use distribution parameters \textit{uncorrected} for broadband magnitude
errors, since these same errors will be present in the photometric
data for the individual
galaxies whose spectra we wish to analyze.

For low-z LRGs, without magnitude error correction, we have 
\begin{eqnarray}
  m^0=&-0.0829 M_V - 0.042  \log_{10}(1+z)& + 0.48 \\
  s^0=&0.006 M_V - 0.09  \log_{10}(1+z)& + 0.22
\end{eqnarray}
and for CMASS galaxies, without magnitude error correction, we have
\begin{eqnarray}
  m^1=&-0.0973 M_V &- 0.60 \log_{10}(1+z) + 0.23 \\
  s^1=&0.0240 M_V &+ 0.76 \log_{10}(1+z)+ 0.49
\end{eqnarray}

We can then take the posterior probability $p(m,s | \{\vec{d}\})$ from the entire
sample as a prior probability for the analysis of an individual
galaxy spectrum.  The posterior probability for $\log_{10} \sigma$ of the spectrum
is then
\begin{equation}
p(\log_{10} \sigma | \vec{d}_i) \propto p(\vec{d}_i | \log_{10} \sigma) \, p(m,s | \{\vec{d}\}) ~.
\end{equation}
Loosely speaking, if the observational error in the velocity dispersion
measured from a single spectrum is comparable to the intrinsic
width $s$ of the particular population from which it is drawn,
then the data and the prior will contribute equally to the determination
of the posterior PDF of $\log_{10} \sigma$.  If the observational error is
small, the effect of the prior will be correspondingly minor, while
if the observational error is large, the posterior PDF will be determined
primarily by the prior.

The application of this method can thus permit a more precise $\sigma$ estimate
for individual galaxies, by making use of
the collective information about the population from which it was drawn.
It is important however to note that if the spectra under consideration are
somehow selected to be biased towards either higher or lower velocity dispersions,
then the prior will pull them systematically towards the population mean,
giving posterior PDFs that are biased relative to the true $\sigma$ values.
We must also be sure only to apply this method to subsamples
of spectra that are much smaller than the population samples used to determine
the distribution parameters.

\section{Discussion \& Conclusion}
\label{sec:disco}

In this paper, we have presented a new technique for estimating
the velocity-dispersion function of LRGs from large numbers of low SNR spectra.
This method incorporates the effects of observational uncertainties
in spectroscopic redshift, velocity dispersion, and broadband magnitude.
We have compared our method favorably to the more traditional
approach of ``stacking'' multiple spectra; our new approach can
perhaps be termed ``Bayesian stacking''.
We have also indicated how the results of our method can be used as
informative priors to provide more precise
estimates of the velocity dispersions of individual galaxies,
provided that those galaxies are an unbiased selection
from the parent distribution at their particular
redshift and luminosity.

We have applied our technique to a sample of 430,000 galaxy spectra from the BOSS project of the SDSS-III,
covering the redshift range from zero to unity, concentrated between approximately $z = 0.2$
and $z = 0.8$.  For the higher-redshift
CMASS target sample (approximately 76\% of
our galaxies), we
detect a highly significant increase in the intrinsic width of the velocity-dispersion
distribution at higher redshifts, indicative of greater galaxy diversity
at fixed luminosity at earlier cosmic times. 
For the lower-redshift LOZ galaxy sample,
we find little evolution in the velocity-dispersion distribution below
$z \approx 0.5$.  Although the CMASS and LOZ samples do not form a
single uniform sample (LOZ galaxies being generally more luminous than CMASS targets
over the range of redshift where the two overlap), our results suggest
that dynamical evolution of massive LRGs is much more significant over the interval
$0.5 < z < 1.0$ as compared to $0 < z < 0.5$.

Future applications of this method to the BOSS galaxy samples
will focus on the effects of observational selection
on the deduced population evolution.  We also plan to divide
our analysis further by rest-frame color, so as to differentiate
between galaxies of different stellar population at a given
redshift and magnitude.  By making a more accurate division
of the sample in terms of stellar mass and star-formation
history, we hope to separate the signatures of dynamical
and stellar-population evolutionary channels, and to thereby
obtain a more detailed picture of LRG population evolution
and a more powerful discriminant between theoretical scenarios.
This approach can also determine whether the effect of increased
population scatter in $\log_{10} \sigma$ at high redshift is due
to greater dynamical diversity, greater stellar-population diversity,
or to some combination of the two effects.



Our measurements can also have important implications for the statistics of gravitational lensing, by
constraining the total lensing cross-section in massive elliptical galaxies between
redshift 0 and 1. Although a precise application to gravitational-lensing
statistics must await a proper treatment of completeness, our current
results can be combined with published luminosity functions \citep[e.g.,][]{cimatti_06,cool_08}
to place a lower limit on the integrated lensing cross section.

The application of hierarchical Bayesian methods such as the one presented here may
hold the key to reconciling the tension between redshift surveys designed for
constraining cosmological parameters and those designed for the study of galaxy evolution.
The former goal generally dictates an SNR just sufficient to measure redshift
for as many galaxies over as large a volume of the universe as possible, while the
latter goal traditionally requires observations at high enough SNR to precisely constrain multiple
physical parameters for each galaxy. However, if the ultimate goal of
galaxy-evolution studies is to measure the
distribution of physical parameters within a statistically significant sample of galaxies, then
Bayesian methods can remove the need to measure those parameters precisely on
a galaxy-by-galaxy basis. In fact, there may indeed be an objective galaxy-evolution
case for trading fewer high-SNR spectra for more low-SNR spectra, so as to
reduce the effects of sample variance. If cosmological experimental
designs can also accommodate the more permissive (e.g., magnitude-limited) targeting
desired for galaxy population studies, then both goals may be well served
by the same redshift survey.

\acknowledgments

The authors thank
David W. Hogg and Glenn van de Ven for valuable
discussion of this work.

Funding for SDSS-III has been provided by the Alfred P. Sloan Foundation, the Participating Institutions, the National Science Foundation, and the U.S. Department of Energy Office of Science. The SDSS-III web site is \texttt{http://www.sdss3.org/}.

SDSS-III is managed by the Astrophysical Research Consortium for the Participating Institutions of the SDSS-III Collaboration including the University of Arizona, the Brazilian Participation Group, Brookhaven National Laboratory, University of Cambridge, University of Florida, the French Participation Group, the German Participation Group, the Instituto de Astrofisica de Canarias, the Michigan State/Notre Dame/JINA Participation Group, Johns Hopkins University, Lawrence Berkeley National Laboratory, Max Planck Institute for Astrophysics, New Mexico State University, New York University, Ohio State University, Pennsylvania State University, University of Portsmouth, Princeton University, the Spanish Participation Group, University of Tokyo, The University of Utah, Vanderbilt University, University of Virginia, University of Washington, and Yale University.

\end{document}